# THREE QUESTIONS ON THE FUTURE OF QUANTUM SCIENCE AND TECHNOLOGY

S. Radenkovic, M. Dugic, I. Radojevic

*Faculty of Science, University of Kragujevac, Radoja Domanovica 12, 34000 Kragujevac, Serbia*

**ABSTRACT.** The answers on the current status and future development of Quantum Science and Technology are presented.

## INTRODUCTION

The Editorial Board of the Kragujevac Journal of Science decided to ask the prominent researchers and scholars to take part in the poll on the status and the future of Quantum Science and Technology. The Poll is stated (but not limited) by the following questions:

- Do you find the Quantum Measurement Problem worth striving?
- Which research directions do you find prominent in the quantum (a) theory and (b) application (including quantum technology)?
- What would you expect of the future post-quantum theory?

Here we thank for all answers received and Ralph Puchta for his initiative and commitment.

## DO YOU FIND THE QUANTUM MEASUREMENT PROBLEM WORTH STRIVING?

*Francesco Buscemi.*[1] I really have no idea if or how the quantum measurement problem may be solved. Quantum theory itself has nothing to say about a putative solution, so, if I must pick one, I would say that any genuine resolution of the measurement problem, if it exists at all, would have to come from new physics for which we currently have neither hints nor any operational need, in a sense. For the moment, the problem seems to tell us more about the limits of our preferred narratives than about the limits of the formalism itself.

*David Deutsch.*[2] It was solved by Hugh Everett in 1957. (And, according to Bitbol (BITBOL, 1996), by Schrödinger a little earlier but not published.)

*Ruth E. Kastner:*[4] Absolutely. The measurement problem is the inability of the theory to specify in physical terms what counts as a "measurement" meriting the assignment of an eigenstate of the measured observable to the system. I have argued that the measurement problem is a serious theoretical anomaly afflicting the conventional formulation of quantum theory, which presupposes that the only actual physical evolution taking place is always unitary, governed by the Schrodinger equation. For example, Everettian approaches remain afflicted by the measurement problem since they cannot differentiate a correlating (entangling) interaction from an interaction yielding a definite outcome, leading to in-principle inconsistency. The latter is revealed in the possibility of empirically detectable discrepancies in Wigner's Friend-type scenarios. (See, e.g., KASTNER, 2020; KASTNER, 2024.)

The conventional adherence to the assumption of unbroken unitary evolution is what produces the measurement problem. It is resolved by adopting a historically present but neglected theory of field behavior, the 'direct action theory' or 'absorber theory of radiation'. My publications, in particular (KASTNER, 2022), present the solution, applicable to the fully relativistic domain, in terms of this theory.

*Olimpia Lombardi*.[5] During the early years after the formulation of quantum mechanics, the orthodox interpretational framework was tied to an instrumentalist reading: quantum features were discussed only in terms of possible measurements results, under the assumption that the concept of quantum measurement should be taken as a primitive concept of the theory. However, in the last decades, the traditional instrumentalist perspective has begun to lose its original strength. On the one hand, from a cosmological viewpoint, a quantum account of the universe cannot be based on measurement understood as an interaction between a system and a measuring apparatus. On the other hand, several realist interpretations of quantum mechanics have been proposed, with the purpose of understanding how reality would be if the theory were true.

From a more general viewpoint, the question is: Why is it worth insisting on the interpretation of quantum physics, not only quantum mechanics but also quantum field theory? If most interpretations are empirically equivalent, what is the point of carrying on the interpretive debates? I am certain that it is worth insisting on the interpretation of quantum physics because only at the interpretative level is a proper understanding of the subject achieved. Of course, the problem now consists in elucidating what is meant by *understanding*.

Understanding is an epistemic state that entails more than mere knowing: I may know that the fall of the Western Roman Empire occurred, but, at the same time, I may not understand that historical process. I consider that this plus of understanding over knowledge involves two features: *projectability* and *transferability*. On the one hand, subjects understand a certain knowledge better to the extent that they can project it further. For example, they can use it successfully to obtain new knowledge, they can infer causal connections from it, they can draw conclusions on its basis, they can transfer it to cases not originally considered and even strongly novel, they can even imagine counterfactual situations that fit it. On the other hand, subjects understand a certain knowledge better to the extent that they can transmit it to other subjects, even those who are not familiar with it. For example, the fact that a scientist can "explain" their topic of study to a layman is a symptom of their understanding, without requiring that the layman achieves a high level of comprehension. Of course, if the topic is very technical, in this process the "explainer" has to appeal to multiple resources, such as the use of images, metaphors, analogies, etc.; but it is precisely their capacity to generate such resources that shows their degree of understanding.

On the basis of these notes of what I conceive as understanding, it is easy to see why I consider that an interpretation of quantum physics provides understanding. If we only possess a formalism with the minimal interpretation that allows theory to be applied, we do not reach an understanding of the matter since we hardly have the necessary tools to project and transfer our knowledge. The full interpretation of formalism is what allows us to extend the quantum knowledge to new situations, to make inferences about which facts are in accordance with the picture provided by the theory, and to evaluate how well the world described by the theory fits into the categories that are at the basis of our traditional conception of reality. And it is only in terms of an interpretation that we can attempt to convey an image of quantum reality through ordinary language.

But even accepting what has been said so far about the notion of understanding, someone of instrumentalist temperament might insist: if the "operational core" of quantum physics has been so overwhelmingly successful in empirical terms, why do we need to understand it through interpretation? My answer is that understanding in the field of quantum physics has consequences for scientific practice itself. Certain interpretations lead to theories that may be more akin to other theoretical domains of physics, such as the conceptual link between quantum mechanics and quantum field theory, or to other scientific disciplines, such as chemistry. Furthermore, the instrumentalist spirit is perhaps one of the factors that discourage young people who seek to understand how the world works and that drive them away from physics. But I believe that the answer to the instrumentalist challenge depends primarily on how each of us conceives of the scientific enterprise. If science is understood exclusively as a means to produce technology in the form of concrete or abstract objects to be injected into consumer society and/or to improve the material lives of human beings, then the discussions about the interpretation of quantum theory make no sense. But those of us who believe that science is an end in itself, because it is essential to the intellectual development of human beings, will continue to think that the understanding provided by the interpretations of quantum physics has intrinsic value.

*Tim Maudlin*.[6] No serious physical theory with pretensions to being exact and fundamental should mention the term "measurement" in its fundamental principles, because that concept is not sufficiently sharp in meaning. "Measurements" are just some vaguely defined class of physical interactions between systems, with no special significance for the basic dynamics. Since some approaches (e.g. the "Copenhagen Interpretation") do employ the concept of measurement, these approaches should be replaced with theories that don't, for example pilot-wave theories or spontaneous collapse theories.

*Philip Pearle*.[7] The short answer is yes! When I took my first course in quantum theory, in my undergraduate Junior year, I saw that there was a problem. The indifference to it of the physics community then (which has been much improved in the 70 years since then) made me think this problem was worth making my life's work. I give a detailed account of my first encounter with quantum theory and my subsequent engagement with the Measurement Problem in the Preface of my recent book (PEARLE, 2024).

Briefly, the Measurement Problem is that, when describing an experiment, the wave function (evolving under Schrodinger's equation) does not correspond to reality, to the single actual outcome seen in the laboratory. Instead, the wave function is the sum of such outcomes, which is pretty weird. But, we need to have the wave function describe reality, so the rules of quantum theory invoke an ill-defined "collapse postulate": the wave function somehow "jumps" to one or another outcome. As I detail in the first chapter of my book, Schrodinger thought this abandonment of his equation was very wrong, and he called it "the most interesting point of the entire theory."

There have been quite a few suggestions for resolving the Measurement Problem, prominent among them being adoption of the deBroglie-Bohm Pilot Wave alternative theory, or choosing some variant of the Many-Worlds theory. My own approach has been one I like to believe Schrodinger would have liked.

I add a term to Schrodinger's equation, so that the wave function smoothly evolves to describe a single outcome of an experiment. The extra term depends on a randomly fluctuating field, and each possible field leads to a definite outcome.

My first paper proposing this resolution was published in 1976, but it wasn't until 1989 that I introduced what I regard as a satisfactory embodiment of this idea, the Continuous Spontaneous Localization (CSL) theory. In the ensuing 36 years, there have been hundreds of theoretical and experimental papers on CSL. The experiments have neither proved CSL is correct nor ruled it out. If it is verified, it means that experiments will show that standard quantum theory is wrong in certain circumstances, and that would be tremendously exciting!

*Lev Vaidman*.[9] Yes.

*Vlatko Vedral*.[10] I think that there is no such thing as the measurement problem in quantum physics. Most of our lack of progress in finding a new theory has to do with the mistaken belief in the existence of the measurement problem. The key to understanding why there is no measurement problem is treating everything quantum mechanically, both the system and the apparatus (or anything else that is relevant for the analysis). I've written extensively about this and the related topics in my upcoming book "Portals to a New Reality" (VEDRAL, 2025).

*Karol Życzkowski*.[11] The "problem of quantum measurement" is an intriguing topic from a philosophical perspective, as ongoing research can deepen our understanding of the foundations of quantum theory. From a physicist's standpoint, I do not expect our current methods for estimating the probability of a particular measurement outcome to change significantly. These calculations, which are essential for experimental work at the microscopic scale, largely remain unaffected by the choice of interpretation of quantum theory. In fact, it is possible to perform complex quantum mechanical calculations and carry out sophisticated experiments without committing to any specific interpretation of quantum theory.

## WHICH RESEARCH DIRECTIONS DO YOU FIND PROMINENT IN (a) THEORY AND (b) APPLICATION (INCLUDING QUANTUM TECHNOLOGY)?

*Francesco Buscemi*. I believe we should more decisively push toward an observer-dependent perspective, not only in quantum theory but in science as a whole, including cosmology. This idea is not new; it is as old as science itself. From time to time it resurfaces, and when it does it sheds new light on physics, yet it is quickly overshadowed again by the powerful illusion that we call "objective reality". In this spirit, I expect research directions that put center stage what an observer can infer and learn, and how such inferences are constrained or enabled by physical theory, to become increasingly prominent. Such an approach may guide both theoretical developments and quantum technologies.

*David Deutsch*. (a) Put quantum field theory on a sound ontological (not just predictive) footing; (b) Not my field.

*Ruth E. Kastner*: There are several prominent directions currently, but the fact of their prominence is not a reflection of their fruitfulness. A general approach of increasing prominence (deservedly so) is exploring the idea that spacetime is not a fundamental background for physical systems but instead is an emergent manifold. One version of this approach, based on the transactional formulation, yields Einstein's equations of the general theory of relativity together with the cosmological constant (replacing the need for 'dark energy') and MOND correction (replacing the need for 'dark matter'); (SCHLATTER, AND KASTNER 2023).

*Olimpia Lombardi*: Nobody can deny the great relevance of quantum technology. Although I am not specialist in technological matters, I think that the most striking practical consequences of quantum theory are quantum information and quantum computing, as well as the production of novel quantum states of matter.

From a theoretical point of view, in addition to specific developments—such as protection against environmental noise or applications of entanglement and non-locality, just to mention a few examples, I believe that the interpretative work should not be overlooked, for the reasons I mentioned in my answer to the previous question. In particular, the research on quantum information and quantum computing has renewed the interest in questions about the foundations of the theory.

*Tim Maudlin*. As far as pure theory, both the pilot wave and spontaneous collapse approaches are worthy of pursuit. Perhaps the Many Worlds theory is, but it still faces basic conceptual issues that have not been resolved. As far as technological design, I have no idea. I am not trained in engineering. There, all one really wants is some easily usable formalism that makes accurate predictions for the sorts of procedures you are interested in as an engineer (e.g. calculation, data processing, information transmission). It is perfectly possible that approaches not sharp enough for foundational theories can serve these purposes.

*Philip Pearle*. Unified with the answer to the third question, please see below.

*Lev Vaidman*: We have to come to agreement about interpretation and continue to look for new quantum devices.

*Vlatko Vedral*. The next two decades of science (and not just physics) will be defined by quantum physics going further and further into the macro domain. Chemistry, Biology, Neuroscience, and so on will ultimately be understood in terms of quantum physics. This will be enabled by the rapid development of quantum technologies, and we will also be witnessing the birth of a large-scale, universal quantum computer.

*Karol Życzkowski*: While it is hardly possible to demonstrate rigorously that quantum theory is the ultimate description of the microscopic world, all experimental data collected so far are consistent with its predictions. As a result, it is likely that quantum theory will continue to be used successfully in describing experiments and predicting new effects throughout the 21st century. Instead of questioning whether quantum mechanics accurately describes micro-scale phenomena, we often focus on identifying which quantum effects can be harnessed for technological purposes.

Recent advances in quantum physics have already led to the development of lasers, diodes, transistors, electron microscopes, and integrated circuits. Furthermore, ongoing experimental progress has rapidly expanded applications such as quantum information processing. Devices for quantum cryptography and quantum communication, which provide theoretically unbreakable security, are already commercially available. However, quantum computing remains in its early stages, with the key challenge being how to achieve a true quantum computational advantage: to solve a concrete and useful computational problem which cannot be solved by the classical computers.

It is expected that this area will continue to evolve rapidly, integrating insights from experimental physics, theoretical physics, mathematics, and information science. Over the next decade, research will likely focus on identifying which quantum states, quantum operations, and quantum protocols will be most useful for quantum information processing.

## WHAT WOULD YOU EXCPECT OF THE FUTURE POST-QUANTUM THEORY?

*Francesco Buscemi*. If a future post-quantum theory ever emerges, to deserve that name, it would have to depart from quantum theory at least as radically as quantum theory departed from classical physics. It is worth recalling that the inadequacy of classical theory was made evident by relatively simple, low-energy experiments at the end of the nineteenth and the beginning of the twentieth century. By contrast, and this is an important methodological point, we currently have no experimentally demonstrated situations in which quantum theory is inadequate. The open issues we do have are conceptual or theoretical rather than empirical. The regimes where quantum theory might conceivably fail lie at extreme scales not yet accessible to experiment. Without concrete empirical hints, imagining a genuinely new framework becomes exceedingly difficult. In this sense, talk of a post-quantum theory today feels uncomfortably close to "armchair philosophizing", carried out without the observational footholds that historically guided genuine theoretical revolutions.

*David Deutsch*. Can't prophesy of course, but I wouldn't be at all surprised if explaining dark energy and/or the inflation field or even dark matter turned out to require post-quantum theories. See also Marletto's constructor-theoretic formulation of thermodynamics (Marletto, 2016), which I consider to be the deepest to date. Not exactly post-quantum yet but certainly complementary to quantum.

*Ruth E. Kastner*: I am not quite sure what is meant by 'post-quantum theory'. Basic quantum theory could well be sufficient once it is reformulated in the direct-action picture, to take into account real physical non-unitary processes, and thus to be able to explain what counts as a "measurement". (See, in particular, Kastner 2022, Chapter 5). And indeed that also gets us a unification of quantum theory and general relativity, as shown in the above publication.

*Olimpia Lombardi*: Of course, we are all waiting for the unification of quantum theory and general relativity. However, regrettably, there is still no far-reaching quantum gravity theory that is both fully articulated and experimentally confirmed, and I have doubts that such a theory will be achieved in the near future.

*Tim Maudlin*. I am not sure what "post-quantum" means here, exactly because there is no agreed-upon understanding of quantum theory. Of course, any future theory must account for the same iconic quantum-mechanical effects—such as two-slit interference, the disappearance of interference when the apparatus includes a monitor, and violations of Bell's Inequality. Regarding the latter, I rather suspect that we may enter a "post-Relativistic regime" rather than a "post-quantum" one. Special and General Relativity are clearly enough defined theories that it will be obvious if a fundamental principle of them is given up. For example, if superluminal signaling can be demonstrated in the lab.

*Philip Pearle*. There is general agreement that the most fundamental problem is to obtain a sound theory that combines General Relativity with Quantum theory. It is my hope that the ideas behind CSL will somehow find a home in this new theory. It may be that this new theory will be quite different from either General Relativity or Quantum Theory but, since both theories are so successful in their respective realms, one would expect that approximations applied to the new theory would yield GR or QT in suitable circumstances.

*Lev Vaidman*: I see no "clouds". There is no reason to look for another theory; quantum predictions and observations fully agree. There will be no post-quantum theory.

*Vlatko Vedral*. I expect the same scientific paradigm to be followed. Our next theory of physics will be a unification and generalisation of quantum physics and general relativity, and it (the next theory) will reduce to quantum physics and general relativity in some special limits. It is impossible to know what this theory will be like, but one thing I can bet on is that it won't be a return to classical physics. Q-numbers are here to stay and, if anything, the fundamental elements of reality in the new theory will only be even weirder.

*Karol Życzkowski*: Post-quantum theories and other *generalized probabilistic theories* (GPT) models will continue to be an active area of research in theoretical physics. Personally, I am interested in this topic and have worked on generalized quantum theories, but I do not believe such theories will have practical significance within my lifetime. It is conceivable that generalized quantum theories could make predictions beyond those of standard quantum mechanics, but these effects are likely limited to extreme conditions—such as very high energy, temperature, density, or acceleration—that are not relevant for the typical experiments of the coming decades.

## GENERAL OPINIONS

*Ivan Gutman*:[3] Since its appearance a little more than a century ago, quantum theory found outstanding, marvelous and magnificent applications, mainly in material science and chemistry, and became by now a standard method of research. Quantum theory yielded results and methods of significant technological importance. Yet, if we would dig a bit deeper, we would see that today Schrödinger's cat is alive as much as it was in 1935. The conundrum envisaged by Einstein Podolsky and Rosen is today as obscure as then, except that we now know that it was not a figment.

*Milena Petković.*[8] The formulation of new chemical concepts and methodologies is expected to enhance our understanding of reaction mechanisms. Significant steps in this direction were taken during the first two decades of the 21$^{st}$ century. For example, analysis of the extreme values of the reaction force enabled a conceptual shift from viewing the transition state as a single structure to recognizing a transition state region, where the key electronic rearrangements associated with bond cleavage and bond formation take place (TORO- LABBÉ, 2009). Further, on the basis of Bader's Atoms in Molecules theory (BADER, 1990), the energy decomposition method known as the Interacting Quantum Atoms approach (BLANCO *et al.*, 2005) was developed; when coupled with the Relative Energy Gradient method, it enables the identification of the driving forces behind chemical transformations (THACKER, *et al.*, 2017). These advances indicate that novel chemical concepts and approaches will continue to emerge, providing even deeper insights into reaction mechanisms.

*[1]Department of Mathematical Informatics, Nagoya University, Furo-cho Chikusa-ku, Nagoya 464-8601, Japan,* E-mail: buscemi@nagoya-u.jp
*[2]Centre for Quantum Computation, the Clarendon Laboratory, Oxford, University of Oxford, United Kingdom,* E-mail: david.deutsch@qubit.org
*[3]Faculty of Science, University of Kragujevac, Radoja Domanovića 12; 34000 Kragujevac, Republic of Serbia,* E-mail: gutman@kg.ac.rs
*[4]Department of Philosophy, University of Maryland, College Park, USA,* E-mail: rkastner@umd.edu
*[5]Department of Philosophy, CONICET, University of Buenos Aires, Buenos Aires, Argentina,* Email: olimpiafilo@gmail.com
*[6]New York University, John Bell Institute for the Foundations of Physics, USA,* E-mail: twm3@nyu.edu
*[7]Physics Department, Hamilton College, Clinton, NY 13323, USA,* E-mail: ppearle@hamilton.edu
*[8]Faculty of Physical Chemistry, University of Belgrade, Studentski trg 12-16, 11158 Belgrade, Republik of Serbia,* E-mail: *milena@ffh.bg.ac.rs*
*[9]The Alex Maguy-Glass Chair in Physics of Complex Systems, Physics Department, Tel Aviv University, Tel Aviv 69978, Israel,* E-mail: lev.vaidman@gmail.com
*[10]Clarendon Laboratory, Department of Physics, University of Oxford, Oxford OX1 3PU, United Kingdom,* E-mail: vlatko.vedral@physics.ox.ac.uk
*[11]Institute of Theoretical Physics, Jagiellonian University (Cracow) and Center for Theoretical Physics, Polish Academy of Sciences (Warsaw), Poland* Email: karol.academicus2@gmail.com